\documentclass[11pt]{article}

\usepackage[final]{acl}

\usepackage{times}
\usepackage{latexsym}
\usepackage{comment} 
\usepackage{ulem}
\usepackage{mdframed}
\usepackage{tabularx}

\usepackage{amsmath}
\useunder{\uline}{\ul}{}
\usepackage{multirow}

\usepackage[T1]{fontenc}

\usepackage[utf8]{inputenc}

\usepackage{microtype}

\usepackage{inconsolata}

\usepackage{graphicx}
\usepackage{subcaption}

\usepackage{footmisc}  
\newcommand{\blfootnote}[1]{%
  \begingroup
    \renewcommand\thefootnote{}
    \footnote{#1}
    \addtocounter{footnote}{-1}
  \endgroup
}

\title{Human Capital Visualization using Speech Amount during Meetings}

\author{Ekai Hashimoto\textsuperscript{1,2}, 
        Takeshi Mizumoto\textsuperscript{2}, 
        Kohei Nagira\textsuperscript{2}, 
        Shun Shiramatsu\textsuperscript{1} \\
        \textsuperscript{1}Nagoya Institute of Technology,Aichi, Japan\\
        \textsuperscript{2}Hylable Inc., Tokyo, Japan\\
        \texttt{e.hashimoto.611@stn.nitech.ac.jp, t.mizumoto@hylable.com}}

\begin{document}
\maketitle

\begin{abstract}
In recent years, many companies have recognized the importance of human resources and are investing in human capital to revitalize their organizations and enhance internal communication, thereby fostering innovation. 
However, conventional quantification methods have mainly focused on readily measurable indicators without addressing the fundamental role of conversations in human capital. 
This study focuses on routine meetings and proposes strategies to visualize human capital by analyzing speech amount during these meetings. 
We employ conversation visualization technology, which operates effectively, to quantify speech.
We then measure differences in speech amount by attributes such as gender and job post, changes in speech amount depending on whether certain participants are present, and correlations between speech amount and continuous attributes. 
To verify the effectiveness of our proposed methods, we analyzed speech amounts by departmental affiliation during weekly meetings at small to medium enterprises.
\end{abstract}

\blfootnote{This paper has been accepted for presentation at
The 26th Annual Meeting of the Special Interest Group on Discourse and Dialogue (SIGDIAL 2025). It represents the author’s version of the work.}

\section{Introduction}
As the economy shifts toward knowledge-intensive value creation, the importance of employees' skills, relationships, and motivations increases. 
In response, many companies invest in their employees, a practice known as Human Capital Management. 
The market also requires measuring the "capital" of employees to assess a company's value.
Furthermore, to facilitate the quantification of human capital, the Japan Cabinet Office released the Human Capital Visualization Guidelines \cite{naikaku}, and ISO standards require companies to disclose quantitative human‑capital information in their annual securities reports \cite{ISO30414}.

Conventional quantification methods for human capital have mainly focused on easy-to-measure indicators such as salaries and the ratio of female managers \cite{Yao_2021, Tomizuka_2017}. 
While these metrics describe a company's state, they offer only indirect insight and are, therefore, insufficient to achieve the underlying goals of human capital visualization. 
In particular, a few indicators related to employee communication have been complex to measure, yet crucial to corporate activities. 
However, we believe conversations play a fundamental role because we inevitably have conversations.

Against this background, the present study focuses on “intra-organizational communication,” an aspect that conventional evaluations have struggled to capture, and proposes a new method that visualizes this communication. 
Specifically, we regard workplace meetings as key settings in which intra-organizational communication occurs and, using the amount of speech for each attribute and its temporal changes, compute (i) the proportion of speech per person for each attribute and its absolute deviation from an ideal distribution, (ii) the difference in the speech amount of one attribute depending on whether another attribute is present, and (iii) the correlation between employee tenure and speech amount.
This study aims to fill this gap by quantitatively visualizing the amount of speech during meetings, enabling a deeper understanding of internal communication patterns and organizational power dynamics.

To verify the effectiveness of the proposed method, we analyzed speech amounts by employee attribute during progress-report meetings at a small-to-medium-sized enterprise.

\section{Related Works}
\subsection{Speech-amount Analysis}
To measure speech amount in meetings quantitatively, this study employs a quantitative discussion analysis technology\cite{Nagira_2023}.
This technology records each participant’s vocal reactions over time and detects the amount of speech—including back-channel cues such as short verbal interjections and laughter—in 5-second segments. 
Non-verbal bodily reactions—e.g., posture changes or head nods—are not captured. 
Because the system reports the ratio of speech amount within a 20-second sliding window, its output ranges from 0 (no speech) to 1 (continuous speech).

The present work does not handle the linguistic content of participants’ utterances.
By excluding content, the method can be applied even to meetings where disclosure of discussion details is undesirable—such as executive sessions or interviews in response to harassment—while still allowing for a quantitative assessment of communication levels.
Accordingly, the proposed approach visualizes disparities in speaking opportunities and speech amounts in a privacy and confidentiality-conscious manner.

However, discussion topics can still affect speech amount. 
Therefore, we divide long-term meeting data into quarterly periods and aggregate each quarter separately. 
This averaging neutralizes topic-specific fluctuations, yielding a more fundamental, instance-independent view of organizational communication.

\subsection{Impact of Participant Attributes on Organizations} 
\citet{ogawa_2024} point out that in Japanese organizational culture, differences in attributes such as gender and job rank affect hesitation to speak.
Variations in speech and communication arising from such attributes are observed in Japan’s hierarchical corporate structures and organizational cultures worldwide \cite{Kalogiannidis_2020}. 
Thus, imbalances across attributes—such as gender, department, and position—are critical factors influencing any organization.

Therefore, this paper aims to capture the balance of communication across organizational attributes more accurately.
To that end, we analyze attribute-specific speech amount in meetings and propose a method that visualizes the power dynamics and inter-attribute relationships embedded in the organization.

\subsection{Inferring Relationships Among Participants} 
Many recent studies have focused on quantitatively estimating the power balance and influence among people engaged in discussions\cite{ijcai2019p645, Hung_2007, otsuka_2006}.  
For example, \citet{Rienks_2006} proposed a model that infers dominance during conversation from speaking order and frequency. 
Most prior studies, however, analyze a single meeting or short-term observations; cases that track temporal changes in speech amount and relationships in dynamic environments such as companies remain limited.

We, therefore, leverage the long-term time-series speech data collected by Hylable Discussion and propose a method for capturing intra-organizational communication.
Specifically, we treat changes and biases in speech amount as indicators of relational values (power relations and degrees of influence) among participants and ultimately visualize the organization’s communication structure over time. 
This, we argue, enables a deeper understanding of attribute-based power relations and makes the evolution of an organization’s human capital observable.



\section{Proposed Methods}
This paper proposes a method to analyze intra-organizational power relations and other attribute-specific dynamics using speech amount data collected during meetings.  
An “attribute” here denotes any group membership, such as gender, job post, or department.

We quantify speech amounts with “Hylable Discussion” and visualize communication for each attribute.  
Specifically, we first collect the speech amount of each participant in a meeting and then aggregate it by attribute before analysis. 
To gauge the impact of HR initiatives, we also treat the long-term time series as explanatory variables to visualize changes in communication patterns.  
Through these efforts, we aim to reveal imbalances and power structures in organizational communication, ultimately providing a new indicator for evaluating human capital.


\subsection{Speech Amount Ratio by Attributes}
We compute the attribute-specific speech-amount ratio to visualize and analyze the communication balance among attributes.
When headcounts differ, attributes with more members naturally exhibit larger total speech. 
Therefore, we divide the total speech of each attribute by its headcount to obtain the per-capita speech amount \(S_a\):

Let \(n_a\) be the number of members belonging to attribute \(a\), and let \(s_{a,i}\) be the speech amount of member \(i\) in attribute \(a\).   
The average speech amount per person \(S_a\)  in attribute \(a\) is

\begin{equation}
  \label{eq: ave_amo_per}
  S_a = \frac{\sum_{i=1}^{n_a} s_{a,i}}{n_a}
\end{equation}

We then define an ideal proportion \(S_a^{\text{ideal}}\) and calculate the absolute error \(E_a\).

\begin{equation}
  \label{eq:ideal_error}
  E_a = \lvert S_a - S_a^{\text{ideal}} \rvert
\end{equation}

The overall error \(E\) is then.

\begin{equation}
  \label{eq:ideal_error_total}
  E = \sum_{a=1}^{n} E_a 
\end{equation}

For example, if equal participation of men and women is desired, one sets \(S_a^{\text{ideal}} = 50\%\) for each.
Other targets, such as encouraging 65\% of female speech and 35\% of male speech, can also be represented.
Although we use equal weights for convenience in this paper, the ideal proportions can be adjusted per a company’s specific objectives.

\subsection{Power Relationships Between Attributes}
In meetings, the presence or absence of a particular department or managerial rank can change the speech amount of other attributes.
For instance, when a department manager attends, staff may speak less, or the presence of the sales team may suppress the development team’s speech.  
Such changes can be analyzed by comparing sessions in which the attribute is present with sessions in which it is absent.

Let \(M_b\) be the set of meetings in which attribute \(b\) is present and \(\bar{M_b}\) the set in which it is absent.   
Let \(l_{M_b}\) denote the number (or total duration) of meetings in each set.
Let \(s_{a, M_b}\) denote the total speech amounts of attribute \(a\) in the meeting \(M\).  

The average speech amount \(S_{a,M_b}\) and \(S_{a,\bar{M_b}}\) of attribute \(a\) when attribute \(b\) is present or absent is shown in \eqref{eq:douseki} and \eqref{eq:un_douseki}.

\begin{equation}
  \label{eq:douseki}
  S_{a,M_b} = \frac{s_{a,M_b}}{l_{M_b}}
\end{equation}
\begin{equation}
  \label{eq:un_douseki}
  S_{a,\bar{M_b}} = \frac{s_{a,\bar{M_b}}}{l_{\bar{M_b}}}
\end{equation}

Define the rate of change in speech amount \(R_{a,b}\)  to the presence of \(b\) on \(a\) as follows.

\begin{equation}
  \label{eq:power_ratio}
  R_{a,b} = \frac{S_{a,M_b}}{S_{a,\bar{M_b}}} - 1
\end{equation}

If \(R_{a,b} < 0\), the presence of attribute \(b\) suppresses the speech of attribute \(a\); conversely, \(R_{a,b} > 0\) indicates that \(a\) speaks more when \(b\) attends.  
Deviation of \(R_{a,b}\) from zero thus suggests latent power structures (authority gradients) among attributes. 
Consequently, deviations from the ideal speech distribution reveal whether the target power balance has been achieved.  
For example, in an organization that seeks a flat hierarchy, a small \(\lvert R_{a,b}\rvert\) when a manager is present versus absent would merit a high evaluation.

\subsection{Continuous-Valued Attributes and Speech Amounts}
When a continuous value can represent an attribute, we can analyze its correlation with speech amount.  
Possible examples include the number of years since hiring or a salesperson’s sales performance.  
Treating such an attribute as an explanatory variable and speech amount as the response allows us to examine behavioral tendencies within the workforce.  

In cultures with a prominent seniority system, such as many Japanese firms, employees with longer tenure may hold more authority.  
By incorporating tenure as an attribute and examining its correlation with speech amount, we can quantify the extent to which a seniority-based communication structure exists within the organization.

\section{Analysis of Real Data Using the Proposed Methods} 
\subsection{Target Data} 
Data collection is logistically challenging because the study relies on Hylable Discussion and requires continuous, long-term recordings.  
Follow-up interviews are also necessary, making large-scale surveys impractical.  
We therefore focused on a single small enterprise for in-depth observation.  

The company has roughly ten employees and holds a weekly one-hour regular meeting.  
All boards and staff attend to share progress on ongoing tasks.  
Each participant presents for three minutes, followed by seven minutes of open Q\&A. 
Because everyone is guaranteed a speaking slot, the setting lends itself to cross-attribute speech comparisons.  
  
Discussion topics naturally cause individual and group-level variance in speech amount.  
While such variance allows micro-level analysis, it hampers macro-level visualization of communication.  
We partitioned the four-year corpus into fiscal quarters (Q) for analysis to smooth topic effects.  
Specifically, after splitting, we summed each participant’s speech amount within every quarter.  
To normalize for unequal meeting counts or lengths, we divided each speech amount by the total duration of meetings that person attended.  

The active headcount ranged from five to ten (Dec 2020 – Oct 2024).  
Quarterly headcount changes appear in Table \ref{tab: human}.  
The corpus comprises a total of 203 meetings.  

Our primary analysis focuses on differences among job positions. 
Gender-based and board/non-board results are given in the appendix because sample sizes were insufficient for robust inference.

\begin{table}[t]
\centering
\small
\setlength{\tabcolsep}{3pt}
    \begin{tabular}{r|cc|cccc}
        Q & \textbf{Male} & \textbf{Female} & \textbf{Board} & \textbf{Sales} & \textbf{Dev.} & \begin{tabular}[r]{@{}l@{}} \textbf{Gen.}\\\textbf{Aff.}\end{tabular} \\ \hline
        2020.4Q & 4 & 3 & 2 & 1 & 3 & 1 \\ 
        2021.1Q & 4 & 3 & 2 & 1 & 3 & 1 \\
             2Q & 4 & 3 & 2 & 1 & 3 & 1 \\
             3Q & 4 & 3 & 2 & 1 & 3 & 1 \\
             4Q & 4 & 3 & 2 & 1 & 3 & 1 \\  
        2022.1Q & 3 & 3 & 2 & 1 & 2 & 1 \\
             2Q & 3 & 3 & 2 & 1 & 2 & 1 \\
             3Q & 3 & 3 & 2 & 1 & 2 & 1 \\
             4Q & 3 & 3 & 2 & 1 & 2 & 1 \\  
        2023.1Q & 3 & 2 & 2 & 1 & 1 & 1 \\
             2Q & 3 & 2 & 2 & 1 & 1 & 1 \\
             3Q & 4 & 2 & 2 & 2 & 1 & 1 \\
             4Q & 4 & 2 & 2 & 2 & 1 & 1 \\  
        2024.1Q & 4 & 2 & 2 & 2 & 1 & 1 \\
             2Q & 6 & 2 & 3 & 2 & 3 & 1 \\
             3Q & 6 & 3 & 3 & 2 & 3 & 1 \\
             4Q & 6 & 2 & 3 & 2 & 3 & 1 \\
    \end{tabular}
\caption{Quarter-by-quarter headcount by attribute (right-most three columns subdivide non-board members).}
\label{tab: human}
\end{table}

\subsection{Speech Ratio for Each Attribute Group}
As shown in Table \ref{tab:percentage_table}, the per-capita speech ratio for each attribute closely matches the ideal 25\%.
Although Sales and General Affairs each constituted only 10\% of the staff until 2023 Q2, they accounted for roughly 23\% of speech per person on average.
Thus, even small groups enjoyed comparable speaking opportunities.

\begin{figure}[!tbp]
  \centering
  \begin{subfigure}{\linewidth}
    \centering
    \includegraphics[width=0.8\linewidth]{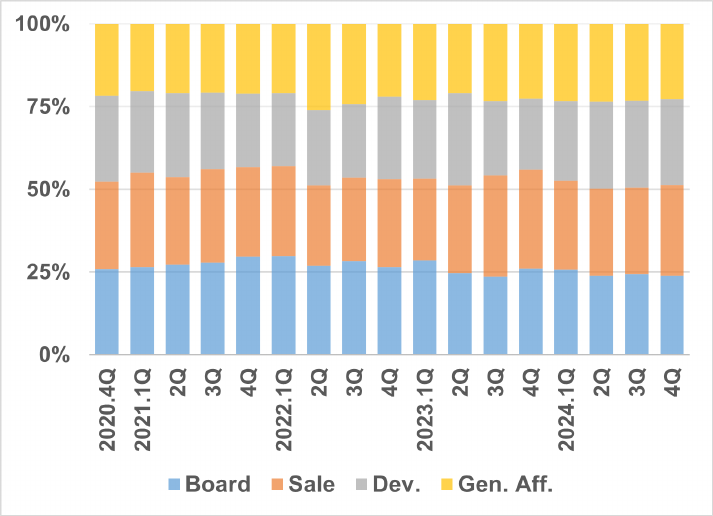}
    \label{fig:sub-a}
  \end{subfigure}
  \hfill
  \begin{subfigure}{\linewidth}
    \centering
    \includegraphics[width=0.8\linewidth]{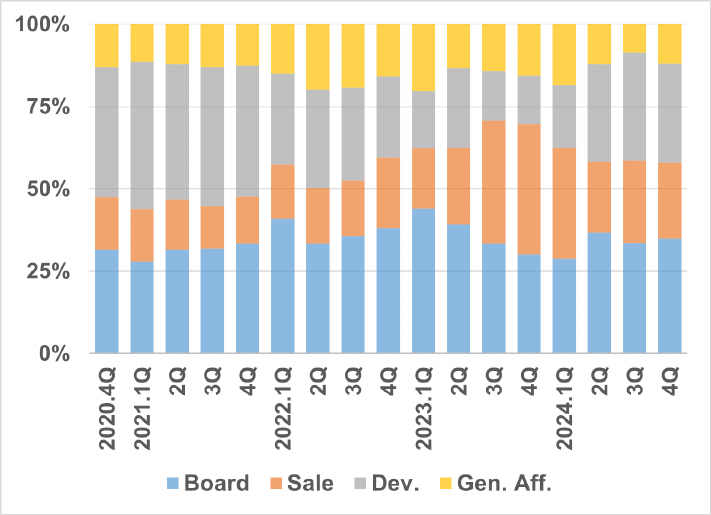}
    \label{fig:sub-b}
  \end{subfigure}
  \caption{Top: speech amount ratio per person by attribute. Bottom: overall speech amount ratio by attribute group.}
  \label{fig:post_speech_amount_ratio}
\end{figure}

\begin{figure}[!tbp]
    \centering
    \includegraphics[width=0.8\linewidth]{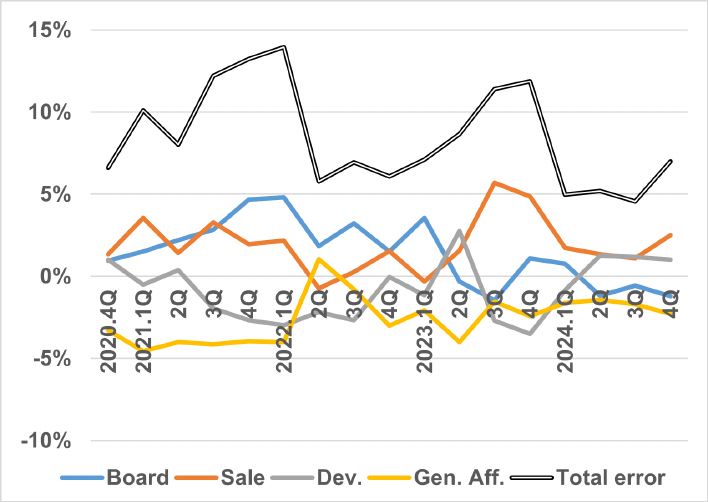}
  \caption{Total absolute error versus ideal 25 \% per attribute.}
    \label{fig:error_total}
\end{figure}

Figure \ref{fig:post_speech_amount_ratio} visualizes the quarterly ratios. 
Figure \ref{fig:error_total} visualizes total error.

From Table~\ref{tab:percentage_table}, Board and Sales dominated from 2021\,1Q through 2022\,1Q.  
This corresponds to intensified sales activity during COVID-19, when boosting revenue became a top priority.

Sales’ share rose again from 2023\,2Q, plausibly because headcount in Sales increased (Table~\ref{tab: human}), temporarily undermining flatness.  
Balance was restored from 2024\,1Q onward.

These results demonstrate how macroeconomic conditions and staffing shifts impact communication.  

Because each participant has a fixed individual report time, speech tends to equalize; less structured meetings (e.g.,  strategy sessions) might display greater attribute-based skew.

\begin{table}[tbp]
\centering
\small
\hspace{-1cm}
\begin{tabular}{r|rrrr|r}
Q & \textbf{Board} & \textbf{Sales} & \textbf{Dev.} & \begin{tabular}[r]{@{}l@{}} \textbf{Gen.}\\\textbf{Aff.}\end{tabular} &\begin{tabular}[r]{@{}l@{}} \textbf{Total}\\\textbf{error}\end{tabular}  \\
\hline
2020.4Q & 1.0\% &  1.3\% &  1.0\% & $-3.3\%$ & 6.6\% \\
2021.1Q & 1.5\% &  3.6\% & $-0.5\%$ & $-4.5\%$ & {\ul 10.1\%} \\
     2Q & 2.2\% &  1.4\% &  0.4\% & $-4.0\%$ & 8.0\% \\
     3Q & 2.8\% &  3.3\% & $-1.9\%$ & $-4.2\%$ & {\ul 12.2\%} \\
     4Q & 4.7\% &  2.0\% & $-2.7\%$ & $-3.9\%$ & {\ul 13.2\%} \\
2022.1Q & 4.8\% &  2.2\% & $-3.0\%$ & $-4.0\%$ & {\ul 13.9\%} \\
     2Q & 1.9\% & $-0.7\%$ & $-2.2\%$ &  1.0\% & 5.8\% \\
     3Q & 3.2\% &  0.3\% & $-2.7\%$ & $-0.8\%$ & 6.9\% \\
     4Q & 1.5\% &  1.5\% &  0.0\% & $-3.0\%$ & 6.1\% \\
2023.1Q & 3.5\% & $-0.3\%$ & $-1.2\%$ & $-2.1\%$ & 7.1\% \\
     2Q & $-0.3\%$ &  1.6\% &  2.8\% & $-4.0\%$ & 8.7\% \\
     3Q & $-1.5\%$ &  5.7\% & $-2.7\%$ & $-1.5\%$ & {\ul 11.4\%} \\
     4Q & 1.1\% &  4.9\% & $-3.5\%$ & $-2.4\%$ & {\ul 11.9\%} \\ 
2024.1Q & 0.8\% &  1.7\% & $-0.9\%$ & $-1.6\%$ & 5.0\% \\
     2Q & $-1.2\%$ &  1.3\% &  1.3\% & $-1.4\%$ & 5.2\% \\
     3Q & $-0.6\%$ &  1.1\% &  1.2\% & $-1.7\%$ & 4.6\% \\
     4Q & $-1.2\%$ &  2.5\% &  1.0\% & $-2.3\%$ & 7.0\% \\
\end{tabular}
\caption{Speech ratio and absolute error by job category (total error \(\ge\)10\% is underlined). Ideal is 25\% }
\label{tab:percentage_table}
\end{table}

\subsection{Power Relationships Between Attributes} \label{sec:power_relation} 
Table \ref{tab:douseki} compares each role’s speech amount when another role is present versus absent.  
Columns indicate the role in attendance; rows show how that presence alters the target role’s speech.  
For example, the Board row / Sales column is --20.8\%, meaning Boards speak ~20.8\% less when Sales is present.  

\begin{table}[tbp]
\centering
\small
\begin{tabular}{l|rrrr}
            & \multicolumn{4}{c}{\textbf{Ratio (present vs absent)}} \\
            & Board & Sales & Gen.~Aff. & Dev. \\ \hline
Board       &  -----            & -20.8\%       & {\ul +20.10\%} & -1.6\% \\
Sales       & {\ul -39.0\%}     & -----         & -21.0\%        & {\ul -42.8\%} \\
Gen.~Aff.   & -2.70\%           & -18.1\%       & -----          & -15.0\%  \\
Dev.        & {\ul -30.7\%}     & {\ul -34.7\%} & -22.2\%        & -----  \\
\end{tabular}
\caption{Change in speech amount when specified attributes are present (0 \%=no change).}
\label{tab:douseki}
\end{table}

Table \ref{tab:douseki} shows that, on average, the presence of another attribute lowers a group’s speech amount by roughly 19.0 points.
Overall, speech declines when other attributes are present, a natural outcome of sharing airtime.

Boards speak 20.1 points more when General Affairs attends; interviews revealed that General Affairs often uses these meetings to gather information from the Board, explaining the increase.

Conversely, Sales and Development speak about 34.9 points less on average when Boards are present, meaning they talk more when Boards are absent, implying substantial influence just below the Board level.

Sales speak far less (–42.8 pt) when Development is present, and Development likewise speaks much less (–34.7 pt) when Sales attend, suggesting that in meetings, each group often takes the lead when the other is absent.

Compared to the presence or absence of Board members, both the Sales and Development teams show a larger decrease in speech amount when the other is present.
This indicates that Boards can participate in either discussion with little change. 
In contrast, Sales and Development, whose domains are more specialized, tend to speak less when their counterparts are present. 
Deepening mutual understanding between Sales and Development would help them converse more freely when both are present.

In short, coexistence / non-coexistence analysis highlights the organization’s communication characteristics and challenges.

\subsection{Continuous Attributes and Speech} 
Here, we quantify the firm’s seniority effect on speech.  
We measured the seniority effect by correlating tenure (the explanatory variable) with quarterly speech amount (the response variable). 

\begin{figure*}[!tbp]
    \centering
    \includegraphics[width=0.8\linewidth]{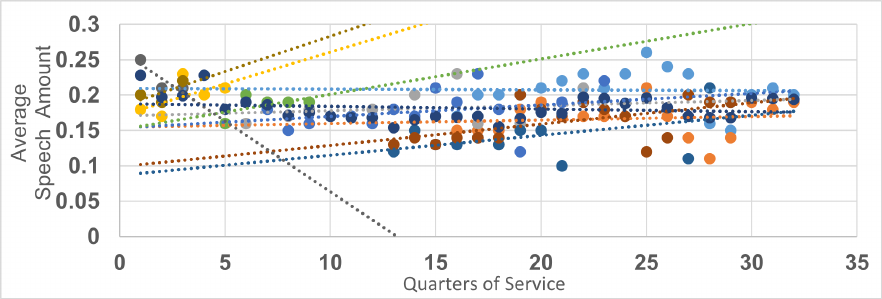}
      \caption{Seniority effect. The vertical axis shows the average amount of speech in each quarter. The horizontal axis indicates a quarter of service. (Each color indicates one person.)}
    \label{fig:nenko}
\end{figure*}

Across all employees, Pearson’s \(r=-0.028\) (Figure \ref{fig:nenko}), essentially none.  
Restricting to those with \> 2 years’ tenure raises \(r\) to 0.560, a moderate positive correlation.  
These findings suggest a positive correlation between employee tenure and speech amount, indicating a seniority-based communication tendency within the organization.

Calculating this metric for each firm or department could reveal cultural differences in seniority-based power structures.

\section{Concluding Remarks}
This paper proposes an analytical method that leverages speech amount data from meetings to visualize a facet of human capital: intra-organizational communication.

Applying the method to approximately four years of weekly progress report meetings at a small enterprise yielded several findings. 
An analysis of speech ratios revealed how the organization’s communication evolved during the COVID-19 pandemic.  
Comparing speech amounts with and without specific attributes clarified the interaction pattern between the board and General Affairs and illuminated the alternating leadership roles of Sales and Development.  
Correlating tenure with speech amount yielded \(r=0.560 \), indicating a moderate positive seniority effect.

The present approach focuses on only the speech amount. 
As a result, the metric could be manipulated, for example, by deliberately increasing one’s speech amount(eg, time and volume) without contributing to human capital.
Future work should combine this quantitative indicator with qualitative analyses of utterance content, enabling a more accurate assessment of human capital.

As \citet{Margariti_2022} demonstrates, providing participants with real-time feedback on their own and others’ cumulative speech amounts reduces imbalances and flattens hierarchical structures.  
Integrating such feedback mechanisms could further advance communication equity and, by extension, improve an organization’s human capital.

\section{Limitations}
This study has several limitations. 

The most significant limitation is the dataset's size. 
Because the target organization is a small, single-employee enterprise, the generalizability of the findings is inevitably limited. 
Future work should collect similar data from multiple, larger organizations to assess the robustness and broader applicability of the proposed approach.

Regarding the validity of our metrics, we conducted only brief, informal interviews with the participating employees, and a systematic evaluation remains outstanding. 
Future work will expand the sample size and cross-check the metrics against objective business indicators to validate the approach further.

Our analysis focuses solely on the speech amount and does not directly assess the significance or constructiveness of individual utterances. 
Consequently, excessive talking could damage the metric; this risk should be considered. 
Future work will explore combining our measure with complementary communication indicators to provide a more balanced assessment.

Building on these limitations, future work should analyze datasets from additional organizations and further validate the metrics to establish a more comprehensive evaluation framework.


\section*{Acknowledgment}
This work was supported by the JST SPRING (JPMJSP2112) and the AIP Challenge Program within the JST CREST (JPMJCR20D1).


\bibliography{custom}


\appendix

\section{Analyzing the Gender}
\begin{figure}[htbp]
  \centering
  \begin{subfigure}{\linewidth}
    \centering
    \includegraphics[width=0.8\linewidth]{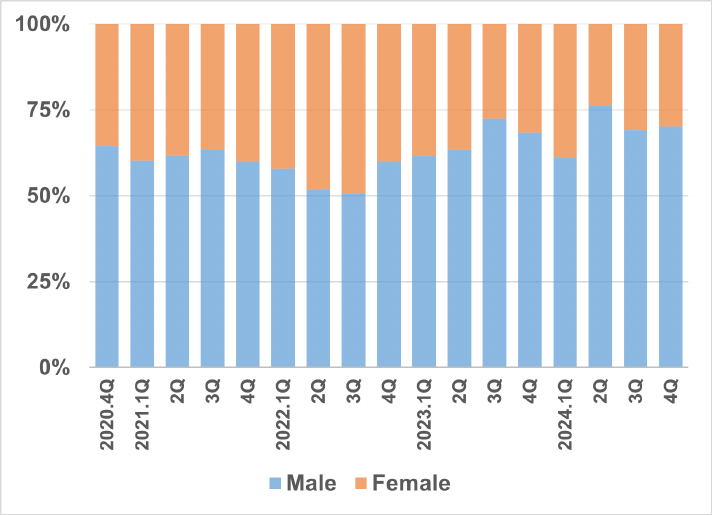}
  \end{subfigure}
  \hfill
  \begin{subfigure}{\linewidth}
    \centering
    \includegraphics[width=0.8\linewidth]{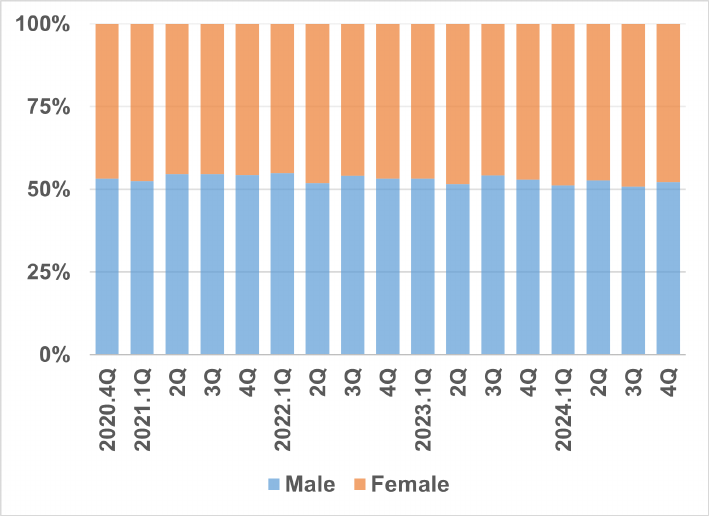}
  \end{subfigure}
  \caption{Top: speech amount ratio per person by gender attribute. Bottom: overall speech amount ratio by gender attribute group.}  
  \label{fig: combined_gen}
\end{figure}

\begin{figure}
    \centering
    \includegraphics[width=0.8\linewidth]{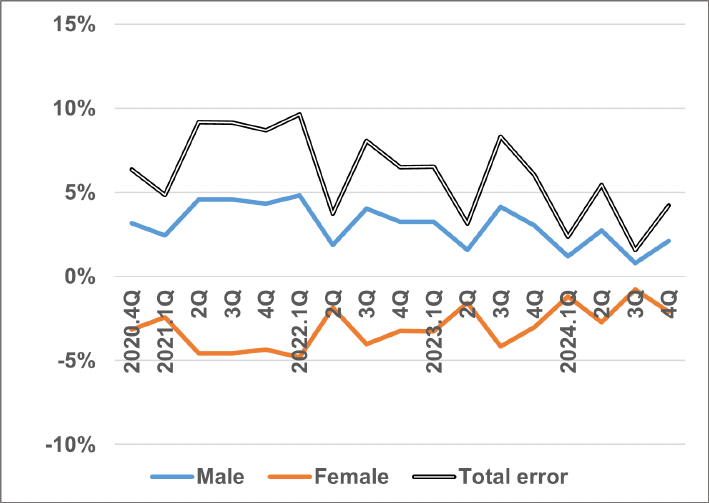}
  \caption{Total absolute error versus ideal 50 \% per attribute.}
    \label{fig:error_total_gen}
\end{figure}

This appendix presents supplementary analyses that focus exclusively on gender.
Figure \ref{fig: combined_gen} traces the per-capita speech ratio for gender across quarters and visualizes the same data in cumulative form.
Table \ref{tab:percentage_table_gen} lists the quarterly per-capita speech ratio for gender and the absolute error from the ideal 50/50 split. 
Figure \ref{fig:error_total_gen} plots the total error series from Table \ref{tab:percentage_table_gen}.
Table \ref{tab:douseki_gen} reports how speech amount changes when the other gender is present.

\clearpage

\begin{table}[!htbp]
\centering
\small
\begin{tabular}{r|rrrr|r}
Q & \textbf{Male} & \textbf{Female} &\begin{tabular}[r]{@{}l@{}} \textbf{Total}\\\textbf{error}\end{tabular}  \\
\hline
2020.4Q & 3.2\% &  $-3.2\%$ & 6.4\% \\
2021.1Q & 2.4\% &  $-2.4\%$ & 4.9\% \\
     2Q & 4.6\% &  $-4.6\%$ & 9.2\% \\
     3Q & 4.6\% &  $-4.6\%$ & 9.1\% \\
     4Q & 4.3\% &  $-4.3\%$ & 8.7\% \\
2022.1Q & 4.8\% &  $-4.8\%$ & 9.6\% \\
     2Q & 1.9\% &  $-1.9\%$ & 3.7\% \\
     3Q & 4.0\% &  $-4.0\%$ & 8.1\% \\
     4Q & 3.3\% &  $-3.3\%$ & 6.5\% \\
2023.1Q & 3.3\% &  $-3.3\%$ & 6.5\% \\
     2Q & 1.7\% &  $-1.7\%$ & 3.1\% \\
     3Q & 4.3\% &  $-4.3\%$ & 8.3\% \\
     4Q & 3.0\% &  $-3.0\%$ & 6.0\% \\ 
2024.1Q & 1.2\% &  $-1.2\%$ & 2.4\% \\
     2Q & 2.7\% &  $-2.7\%$ & 5.5\% \\
     3Q & 0.8\% &  $-0.8\%$ & 1.6\% \\
     4Q & 2.1\% &  $-2.1\%$ & 4.2\% \\
\end{tabular}
\caption{Speech ratio and absolute error by gender. Ideal is 50\%}
\label{tab:percentage_table_gen}
\end{table}

\begin{table}[!htbp]
\centering
\small
\begin{tabular}{l|rr}
\multicolumn{3}{c}{\textbf{Ratio (present / absent)}} \\
        & Male & Female \\ \hline 
Male    & ---                     &  $-43.7\% $  \\
Female  & N/A\textsuperscript{*}  & ---           \\
\end{tabular} \\
\caption{Change in speech amount when attributes are present vs.\ absent. ({*} Division by zero: no absent group to compare with.)} 
\label{tab:douseki_gen}
\end{table}

\newpage

\section{Analyzing on Board vs Non-board}
\begin{figure}[htbp]
  \centering
  \begin{subfigure}{\linewidth}
    \centering
    \includegraphics[width=0.8\linewidth]{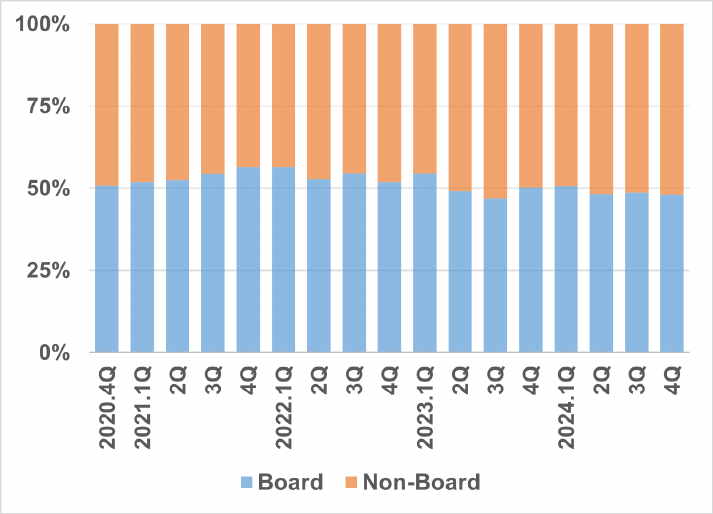}
  \end{subfigure}
  \hfill
  \begin{subfigure}{\linewidth}
    \centering
    \includegraphics[width=0.8\linewidth]{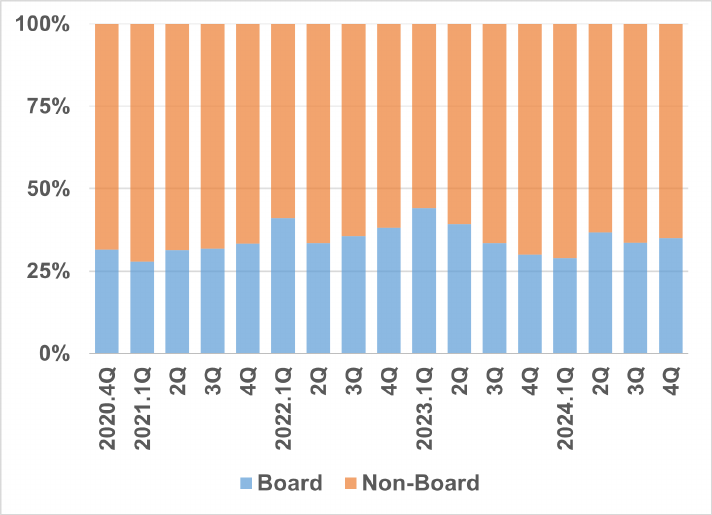}
  \end{subfigure}
  \caption{Top: speech amount ratio per person by attribute. Bottom: overall speech amount ratio by attribute group.}  
  \label{fig: combined_non_board}
\end{figure}

\begin{figure}
    \centering
    \includegraphics[width=0.8\linewidth]{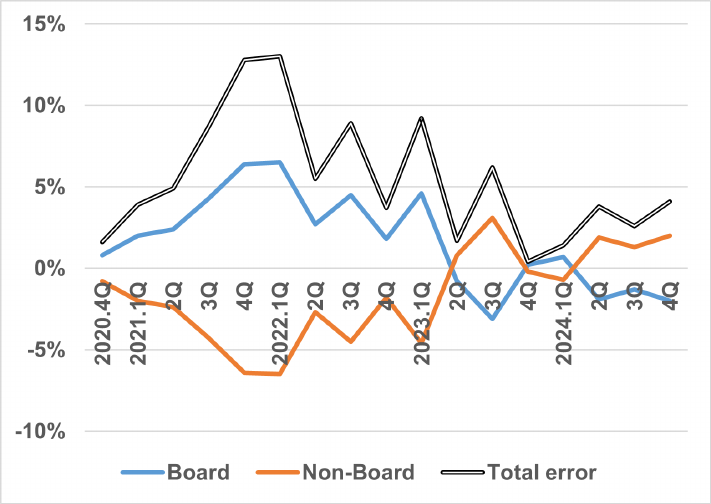}
  \caption{Total absolute error versus ideal 50 \% per attribute.}
    \label{fig:total_non}
\end{figure}

This appendix provides supplementary analyses contrasting board members with non-board employees.
Figure \ref{fig: combined_non_board} traces the per-capita speech ratio for the board and non-board across quarters and visualizes the same data in cumulative form.
Table \ref{tab:percentage_table_non} reports the per-capita speech ratio for board and non-board and the absolute error from the ideal 50/50 split. 
Figure \ref{fig:total_non} graphs the total-error series from Table \ref{tab:percentage_table_non}. 
Table \ref{tab:douseki_non} shows how speech amount changes when the other group is present. 

\begin{table}[!htbp]
\centering
\small
\begin{tabular}{r|rrrr|r}
Q & \textbf{Board } & \textbf{Non-board} &\begin{tabular}[r]{@{}l@{}} \textbf{Total}\\\textbf{error}\end{tabular}  \\
\hline
2020.4Q & 0.8\%     &  $-0.8\%$ & 1.6\% \\
2021.1Q & 2.0\%     &  $-2.0\%$ & 3.9\% \\
     2Q & 2.4\%     &  $-2.4\%$ & 4.9\% \\
     3Q & 4.3\%     &  $-4.3\%$ & 8.7\% \\
     4Q & 6.4\%     &  $-6.4\%$ & 12.8\% \\
2022.1Q & 6.5\%     &  $-6.5\%$ & 13.0\% \\
     2Q & 2.7\%     &  $-2.9\%$ & 5.5\% \\
     3Q & 4.5\%     &  $-4.5\%$ & 8.9\% \\
     4Q & 1.8\%     &  $-1.8\%$ & 3.7\% \\
2023.1Q & 4.6\%     &  $-4.6\%$ & 6.5\% \\
     2Q & $-0.8\%$  &  -0.8\% & 1.7\% \\
     3Q & $-3.1\%$  &  3.1\% & 6.2\% \\
     4Q & 0.2\%     &  $-0.2\%$ & 0.4\% \\ 
2024.1Q & 0.7\%     &  $-0.7\%$ & 1.4\% \\
     2Q & $-1.9\%$  &  -1.9\% & 3.8\% \\
     3Q & $-1.3\%$  &  -1.3\% & 2.6\% \\
     4Q & $-2.0\%$  &  2.0\% & 4.1\% \\
\end{tabular}
\caption{Speech ratio and absolute error by job title. Ideal is 50\%}
\label{tab:percentage_table_non}
\end{table}

\begin{table}[!htbp]
\centering
\small
\begin{tabular}{l|rr}
\multicolumn{3}{c}{\textbf{Ratio (present / absent)}} \\
        & Board & Non-board \\ \hline 
Board    & ---                     &  N/A\textsuperscript{*}  \\
Non-board  & -29.8\%  & ---           \\
\end{tabular} \\
\caption{Change in speech amount when attributes are present vs.\ absent. ({*} Division by zero: no absent group to compare with.)} 
\label{tab:douseki_non}
\end{table}
\end{document}